\pdfoutput=1
\documentclass[manuscript,nonacm]{acmart}
\usepackage{graphicx,booktabs,tabularx,array,needspace,placeins}
\begin{document}
\raggedbottom
\title[Chameleon: Exploring Risk Interpretation in VR]{Chameleon: Exploring Risk Interpretation through Scene-Matched Textures in Virtual Reality}
\author{Yichen Andy Yu}
\affiliation{%
  \department{Department of Computer Science}
  \institution{North Carolina State University}
  \city{Raleigh}
  \state{North Carolina}
  \country{United States}}
\email{yyu55@ncsu.edu}

\author{Wanru Jovie Li}
\affiliation{%
  \institution{Carnegie Mellon University}
  \city{Pittsburgh}
  \state{Pennsylvania}
  \country{United States}}
\email{wanrul@alumni.cmu.edu}

\author{Yang Zhan}
\affiliation{%
  \department{Computer Science}
  \institution{North Carolina State University}
  \city{Raleigh}
  \state{North Carolina}
  \country{United States}}
\email{yzhan3@ncsu.edu}

\author{Yuxuan Huang}
\affiliation{%
  \department{Computer Science}
  \institution{North Carolina State University}
  \city{Raleigh}
  \state{North Carolina}
  \country{United States}}

\author{Mandy Lui}
\affiliation{%
  \institution{University of Rochester}
  \city{Rochester}
  \state{New York}
  \country{United States}}
\email{mlui@u.rochester.edu}

\author{Yiqi Guo}
\affiliation{%
  \department{Department of Computer Science}
  \institution{North Carolina State University}
  \city{Raleigh}
  \state{North Carolina}
  \country{United States}}
\email{yguo36@ncsu.edu}

\author{Qiao Jin}
\affiliation{%
  \institution{North Carolina State University}
  \city{Raleigh}
  \state{North Carolina}
  \country{United States}}

\renewcommand{\shortauthors}{Yu et al.}
\begin{abstract}
Users in virtual reality need to judge contact risks around physical objects while remaining engaged in the virtual scene. Scene-matched textures give physical objects a place in virtual reality and provide visual cues for interaction. We examine how Chameleon's risk-aware textures are perceived during a reaching task. A study with 42 participants compared scanned appearances with red warnings, uniform environment textures, and risk-aware scene-matched textures. Participants rated the risk-aware condition more favorably on four aspects of perceived cue clarity than both alternatives. Interviews connected visible warnings with initial caution and linked confidence in physical interaction to material expectations and intended actions. Awareness of an object's location did not necessarily imply confidence in interacting with it. These findings motivate risk cues that connect depicted materials to the physical properties relevant to an intended action, and ways to support verification during interaction.
\end{abstract}

\begin{teaserfigure}
  \centering
  \includegraphics[width=\textwidth]{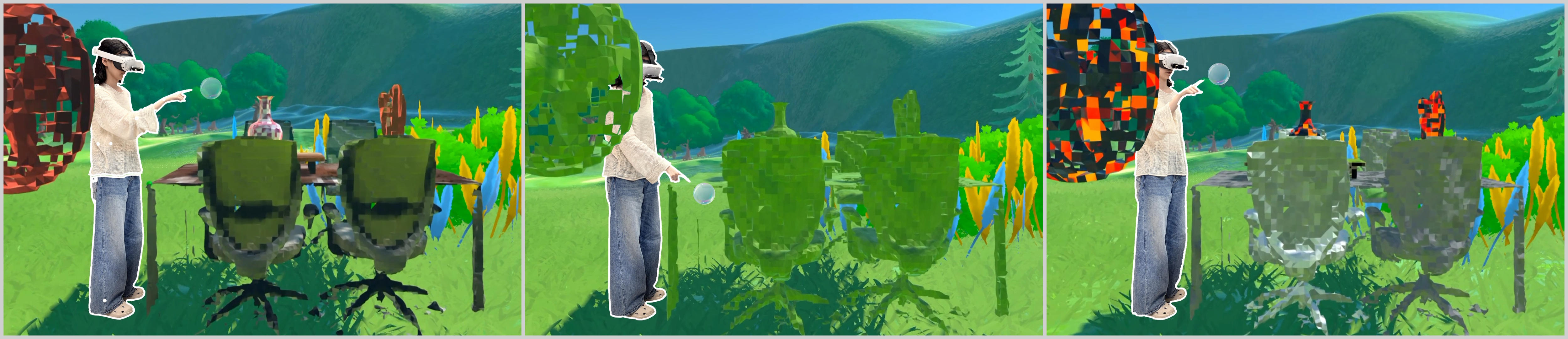}
  \caption{System screenshots of the three rendering conditions. Left: Baseline preserves scanned object textures and adds a red-light cue to high-risk objects. Middle: Condition 1 (C1) applies a uniform environment texture. Right: Condition 2 (C2) applies scene-matched textures differentiated by assigned risk level.}
\Description{Three side-by-side screenshots compare reconstructed physical objects in a woodland virtual environment. The left view retains the objects' scanned appearance with red emphasis on high-risk objects. The middle view gives the objects a shared environment texture. The right view uses different scene-matched textures to convey assigned risk levels.}
  \label{fig:teaser}
\end{teaserfigure}

\maketitle
\section{Introduction}
Room-scale virtual reality (VR) enables users to explore virtual environments through physical movement and interaction~\cite{chameleon_A01,chameleon_C12,Murillo2017NaviFields,Zhang2020Virtual}. Physical objects can also provide surfaces for touch, extending what users see into what they can feel~\cite{chameleon_B04,chameleon_B09,chameleon_B13,chameleon_B05,Abtahi2019Beyond}. These opportunities make the representation of nearby objects important: users need to understand where a physical surface is and how to approach or touch it while remaining engaged in the virtual scene. The design challenge is therefore to integrate physical objects into VR while communicating the degree of caution appropriate for interacting with them.

Existing approaches offer several ways to connect virtual experiences with physical surroundings. Showing physical objects can support awareness of the room~\cite{chameleon_A02,chameleon_B08}, while substitutional reality gives those objects appearances suited to a virtual world~\cite{chameleon_B01,chameleon_D01}. Work on cross-reality interaction similarly explores how physical objects can become resources for virtual activities~\cite{chameleon_C08,chameleon_C02,Lyu2026Unbounded,Lyu2026Objestures}. Scene adaptation and authoring approaches extend this correspondence to the surrounding space, arranging or transforming virtual content in relation to the room~\cite{Gai2017Supporting,Marwecki2018Scenograph,chameleon_C15,chameleon_D03,Yue2017SceneCtrl}.

Visual integration also changes the information users have about physical contact. A substituted material can suggest properties such as hardness or stability, making the relationship between appearance and touch relevant to interaction~\cite{chameleon_B01,chameleon_B11,chameleon_D06}. Applying one environment texture to different objects may make them look coherent while obscuring distinctions between soft and rigid surfaces. A conspicuous warning can direct attention to an object that warrants caution, but may disrupt immersion if its appearance conflicts with the virtual scene~\cite{Lange2020HiveFive}. This raises a question about combining these goals: how can scene-matched surfaces communicate differences in risk during interaction? Evaluating such cues requires considering users' risk judgments alongside their experience of the scene and their activity around physical objects.

To investigate this question, we conducted a between-participants study with 42 participants, comparing scanned appearance with a high-risk red-light cue (Baseline), uniform environment-texture substitution (C1), and scene-matched, risk-aware textures (C2). Participants completed a bubble-reaching task under one condition. Questionnaire responses assessed perceived risk communication and experience in VR; reconstructed contact and clearance measures characterized activity around objects. Semi-structured interviews explored how participants interpreted the representations and made decisions about physical interaction.

Chameleon provided the risk-aware rendering used in this comparison, applying scene-matched textures to physical objects reconstructed through live scanning. The scanned meshes preserve reconstructed contours and positions, while material appearance, color, and contrast jointly express three assigned categories of caution around contact. Our comparison examines how users perceive and act around objects under these complete rendering strategies.

C2 received higher ratings for four aspects of perceived cue clarity than both alternatives and higher self-reported immersion ratings than Baseline. Participants in C2 also showed greater estimated clearance from high-risk surfaces than both alternatives and fewer estimated high-risk contact episodes than C1. Interview accounts connected visible warnings with initial caution and showed how confidence depended on the intended use of a surface: locating a table's boundary and judging whether to lean on it involved different physical expectations.

This paper makes three contributions:
\begin{itemize}
  \item An empirical comparison of three complete rendering strategies in terms of perceived cue clarity, experience in VR, and estimated contact and clearance around physical objects.
  \item An account of how warning cues, material expectations, and physical confirmation inform judgments about contact and the intended use of a surface.
  \item Design implications for relating scene-matched risk cues to contact-relevant properties and supporting verification during interaction.
\end{itemize}

\section{Related Work}\label{sec:related}
Representing physical surroundings in VR allows users to remain aware of nearby objects and incorporate them into virtual activities. How those objects appear also shapes expectations about interacting with them. Our work draws on research into physical awareness, physical-to-virtual substitution, and scene-integrated cues to investigate how surface appearance can communicate different degrees of caution.

\subsection{Awareness of Physical Surroundings in VR}
Physical-world awareness is a continuing concern in room-scale VR because users move through a physical room while attending to virtual content~\cite{chameleon_A03,Marwecki2018VirtualSpace,Azmandian2017An}. Limited access to the surroundings can interfere with everyday activities such as using a keyboard, taking a drink, or noticing another person~\cite{chameleon_A02}. Recorded breakdowns also include collisions with furniture and spectators~\cite{chameleon_A06}. Visible safety boundaries provide a spatial reference, but their interpretation depends on users' understanding of the room. Observed boundary crossings include deliberate decisions to continue an activity based on knowledge of the physical surroundings~\cite{chameleon_A05}. Awareness therefore involves interpreting available spatial information and deciding how to act on it~\cite{Xu2020Exploring,Sukan2014ParaFrustum}.

Selective access to physical information can support task performance and presence during interaction with real objects~\cite{chameleon_A02}. Representations ranging from 3D reconstructions to avatars and images make nearby people and objects visible within VR~\cite{chameleon_B08,chameleon_A11,chameleon_A14,chameleon_A10}, while customizable views control where, when, and how much reality is shown~\cite{chameleon_C07,chameleon_C02}. The appropriate presentation also depends on who needs the information~\cite{Wang2020Slice,Krug2026Mixed}. Proxemic cues use distance, orientation, and location to support awareness of nearby people~\cite{chameleon_A13,chameleon_A15}. Across the headset boundary, alerts inform bystanders about the guardian area, while shared-space tools let them communicate their own spatial needs~\cite{chameleon_A04,Yang2018ShareSpace}. Preferences also reflect the relationship between the people involved, the demands of an interruption, and the user's desired immersion~\cite{chameleon_A08,chameleon_A16}. Awareness cues therefore need to fit the social and spatial circumstances of an activity~\cite{Zhang2025Do,Isaza2024Understanding,Lu2026MoveTogether}.

The experience of moving around physical objects includes concerns about possible collisions and difficulties with orientation, as captured by the Collision Anxiety Questionnaire~\cite{chameleon_A22}. Encounters with obstacles can also affect spatial presence: estimates from a study of physical-environment knowledge associated repeated obstacle avoidance with lower presence~\cite{chameleon_C10}. These findings make users' experience of the surrounding space relevant alongside the visibility of its contents. For an object already represented in VR, a further question is how its appearance can help users judge the caution associated with approaching it. This relationship between physical awareness and perceived caution motivates comparing how users interpret different representations of the same physical objects.

\subsection{Physical-to-Virtual Substitution and Material Expectations}
Substitutional reality incorporates the physical room into a virtual experience by pairing real objects with virtual counterparts, including through workflows for authoring playable environments~\cite{chameleon_B01,chameleon_C11}. Reconstruction-based generation uses detected obstacles and walkable areas to shape virtual spaces for individual and shared use~\cite{chameleon_B03,chameleon_C01}. Correspondence can also be maintained during movement~\cite{Thomas2020Towards}: procedural generation in unknown buildings adapts virtual objects to newly detected obstacles~\cite{chameleon_B06}, while outdoor walking systems align physical routes with virtual paths and update the scene around encountered obstacles~\cite{chameleon_B07}. At the room scale, virtual obstacles with comparable spatial occupancy have been evaluated against conventional chaperone displays~\cite{chameleon_A18}. These approaches use physical geometry to constrain scene construction~\cite{chameleon_C14,Han2022Foldable} and establish the surfaces users encounter.

Represented objects also create expectations about contact~\cite{Fang2020Wireality}. Research on haptic proxies considers both the fidelity of tangible objects and their correspondence with virtual content~\cite{chameleon_B10,chameleon_B12,chameleon_B18,chameleon_D05}. Everyday objects can be selected for a desired interaction~\cite{chameleon_B02,Cheng2018iTurk,Wang2020MoveVR}, while visual remapping lets one physical prop support multiple virtual objects or contact locations~\cite{chameleon_B13,chameleon_B14,chameleon_B15,chameleon_B16,chameleon_B20,Fang2021Retargeted}. Perceptual evaluations examine motion discrepancies~\cite{Zenner2019Estimating,Hoshikawa2024RedirectedDoors} across grasp and manipulation conditions~\cite{Feick2022Designing,chameleon_B19}, as well as differences in apparent object size~\cite{Bergstrm2019Resized,chameleon_B21}. Haptic fidelity also involves consistency between touch and other sensory information~\cite{chameleon_B11}. In generated haptics, inferred virtual material properties guide the vibration assigned to scene objects~\cite{chameleon_D04}. Together, this work makes material expectations relevant to representation. A stone-like texture may suggest rigidity or weight, while vegetation may suggest softness; using these associations to communicate caution requires accounting for these contact expectations.

Functional and contextual correspondence provide further criteria for relating virtual content to the room~\cite{Li2026RealTwin,Jin2026BehaviorAware,Liang2021SceneAware}. Authored mappings connect content to physical interaction affordances and semantic relationships~\cite{Lin2020ARchitect,chameleon_C04,Wang2020CAPturAR,Suzuki2020RealitySketch}. Adaptive layouts likewise use semantic object associations and available input surfaces to position content~\cite{chameleon_C03,chameleon_C05,Kwon2026Draped}. Pose-aware substitution preserves the behavior of people and vehicles while giving them identities suited to another setting~\cite{chameleon_C06}. In Roomify, scanned furniture receives themed counterparts designed to preserve geometric and functional properties. Its mapping uses a binary flag for objects likely to be contacted and requires users to confirm their placement before entering VR~\cite{chameleon_D01}. These approaches connect scene design to the properties and uses of the physical environment. Our comparison examines how different appearances of reconstructed objects shape users' material expectations and judgments about physical interaction.

\subsection{Communicating Caution through Scene-Integrated Cues}
Embedding physical events within the virtual setting can help preserve the continuity of an experience~\cite{Kari2025Reality}. Diegetic representations of an interrupting person supported a stronger sense of place and less disruptive interactions during a virtual office task~\cite{chameleon_A09}. Physical distractions can acquire an explanation within the scene, such as real airflow corresponding to moving virtual trees~\cite{Tao2022Integrating}; integrating pet activity as interactive objects also improved ratings of realism and enjoyment over a baseline without pet integration~\cite{chameleon_A20}. The benefits depend on the representation and setting. Comparisons of scene-matched avatars and abstract obstacle displays in outdoor VR found differences in effort, frustration, and preference~\cite{chameleon_A21}. Scene integration therefore calls for examining how a representation fits an activity and how users experience it.

Scene integration also concerns how information attracts attention~\cite{Kim2025On,Lai2026Searching,Marwecki2019MiseUnseen}. Directional guidance through swarm motion combined fast responses with comparatively little disruption to immersion~\cite{Lange2020HiveFive}. A moving gap in a safety-boundary wall reduced detection time for approaching bystanders compared with Halo and Radar cues~\cite{chameleon_A07}. These examples connect noticeability to a visual event that users can interpret within the scene. For persistent object representations, the corresponding design concern is whether a surface can remain consistent with its surroundings while making greater caution sufficiently conspicuous.

Caution can be encoded in a cue's visual form and adapted to the current interaction~\cite{Jung2018Ensuring,Kang2019Obstacle}. Green and red overlays distinguish grabbable and hazardous tool parts for people with low vision~\cite{Lee2024CookAR}, while yellow, orange, and red volumes communicate three hazard levels around a robot during collaboration~\cite{SanMartin2025HazardZones}. Risk assessment can also govern other content: combining attention monitoring with scene-based hazard assessment enables predicted risk levels to control augmented reality (AR) overlay transparency and warning cues~\cite{Pei2025AttentionAR}. These approaches associate visual differences with the action or degree of caution relevant to a physical situation.

More broadly, adaptive mixed reality (MR) interfaces use cognitive load, task requirements, and environmental or social cues to govern content placement and detail~\cite{Lindlbauer2019ContextAware,chameleon_C13}. Selective suppression and abstract proxies offer ways to reduce visual information or simplify physical-object selection~\cite{Cheng2022Towards,Liu2025Reality}. Interaction-state blending selects virtual proxies, affordance contours, passthrough, or boundary boxes as users perceive, approach, manipulate, or avoid an object~\cite{Seo2024GradualReality}. This work connects representation to the information needed for an action. Our study examines how users interpret apparent materials as cues for caution during movement and reaching.

Representation can affect movement and contact as well as subjective experience~\cite{chameleon_A10,Rietzler2018Rethinking,Nagao2018Ascending}. Virtual-obstacle width, disappearance effects, and multimodal wall feedback have altered whether users pass through or avoid virtual constraints~\cite{chameleon_D02,Boldt2018You}. In assistive AR, mapping depth to a high-contrast color overlay reduced collisions during mobility and grasping tasks among participants with retinitis pigmentosa~\cite{Angelopoulos2019EnhancedDepth}. These findings motivate examining activity around represented objects alongside impressions of the display. Our evaluation therefore considers contact estimates and closest approach by object risk together with reports of risk communication and experience in VR.

Our study compares perceived risk communication, scene experience, and activity around objects under three rendering strategies during movement and reaching. Our contribution concerns how users interpret these cues: how an appearance signals caution, what physical properties it suggests, and how users judge a particular action.

\section{Chameleon}\label{sec:system}
Chameleon represents physical objects as parts of a virtual scene and uses their surface appearance as an intended cue to different degrees of caution. Texture generation takes the object's assigned risk level and a screenshot of the current virtual environment as inputs. A user moving or reaching through the scene encounters the object's reconstructed shape with a texture generated to convey that level within the scene. Geometry provides a spatial reference; apparent material and visual prominence provide cues whose interpretation we examine in the study.

\subsection{Design Rationale}
Chameleon's rendering design follows three priorities: preserving a spatial reference to physical objects, keeping material expectations relevant to contact, and making greater caution visible within the scene.

\paragraph{Preserve the object's spatial reference.}
Chameleon uses the scanned mesh as the surface for visual substitution. Changing its texture preserves the reconstructed contour and placement, giving the user a spatial reference while choosing where to walk or reach. The reconstruction supplies the geometric correspondence between the physical object and its virtual representation~\cite{Nuernberger2016SnapToReality,Li2026RealTwin}.

\paragraph{Keep material expectations relevant to contact.}
A substituted material can imply weight and mobility as well as hardness~\cite{chameleon_B01}. We therefore use material character to express the care associated with contact, with attention to the additional actions an appearance may suggest. For example, incidental contact with a soft-looking surface and using that surface for support involve different expectations about its physical properties.

\paragraph{Make greater caution visible within the scene.}
A recognizable motif needs enough visual prominence to be noticed during an activity~\cite{Kim2025On,Lai2026Searching}. Texture generation consequently combines scene-compatible material associations with differences in contrast and brightness: lower-risk surfaces share the surrounding palette, while higher-risk surfaces receive more conspicuous treatment. Scene coherence and prominence serve complementary purposes in this design.

\subsection{User Experience and Interaction}
Users encounter the physical room through its reconstructed surfaces embedded in the current virtual environment. These surfaces occupy their scanned locations and receive textures that fit the scene while conveying low, medium, or high assigned risk. The cue appears on the object to which it refers, so users can inspect its shape and appearance together while moving through the scene.

In the woodland environment used for our study, consider a user reaching for a floating bubble near a rigid object. The object's contour indicates the space occupied by the physical surface; a bark-like texture is intended to suggest a surface that should be approached with care. Soft vegetation offers a different intended contact expectation, while a flame or thorn motif is intended to suggest keeping clear. The study examines how users interpret these distinctions during interaction.

Chameleon provides the visual information on the surrounding surfaces throughout the activity, allowing attention to move between the virtual target and the physical space around it. The study in Section~\ref{sec:study} examines this interaction during a task that repeatedly involves movement and reaching.

\subsection{Representing Risk through Scene-Matched Textures}\label{sec:vocabulary}
Children's furniture safety guidance concerning sharp edges, electrical components, and mechanisms~\cite{Bulbul2018Hazards} informs the physical concerns considered in our risk classification. We use low, medium, and high risk as operational categories for the caution an object representation is intended to convey during movement and reaching. Low covers soft objects without sharp features, such as bean bags, for which incidental contact calls for relatively little caution. Medium covers rigid objects, such as tables and chairs, that permit contact and interaction with care. High covers objects with designated contact hazards, including heat-generating appliances such as ovens and microwaves, sharp features such as those on some hangers, and moving mechanisms such as those in fans. The distinction is between relatively forgiving contact, contact requiring care, and contact to avoid; these labels express intended contact guidance rather than estimates of injury probability or severity.

The assigned category specifies the caution to convey, while a screenshot of the current virtual environment provides the visual context for texture generation. The model infers the scene's theme, palette, and visual style from the screenshot, then generates a texture whose material appearance, color, and contrast express the assigned category within that setting.

Our study used a woodland virtual environment, so the scene screenshot guided generation toward woodland-related appearances. Table~\ref{tab:vocabulary} provides selected examples of material associations and intended contact cues for this setting. The examples are non-exhaustive; generated appearances depend on the scene screenshot and assigned risk level.

\begin{table*}[t]
\caption{Illustrative, non-exhaustive examples of texture motifs and intended contact interpretations for each assigned risk level in the study's woodland environment.}
\label{tab:vocabulary}
\small
\begin{tabularx}{\textwidth}{@{}p{0.08\textwidth}p{0.24\textwidth}Xp{0.25\textwidth}@{}}
\toprule
Level & Contact concern & Example texture motifs & Intended interpretation \\
\midrule
Low & Soft surfaces without sharp protrusions & Moss, grass, or bushes; subdued color and a yielding appearance & Relatively little caution around incidental contact \\
Medium & Rigid surfaces that require care on contact & Bark, vine, or stone motifs; greater contrast and a harder appearance & Notice the surface and approach with care \\
High & Potential heat, sharp features, or moving mechanisms & Flame or thorn motifs; conspicuous brightness or contrast & Keep clear of the represented object \\
\bottomrule
\end{tabularx}
\Description{Three rows give non-exhaustive examples of texture motifs for the study's woodland environment: soft vegetation for low assigned risk, harder natural materials for medium assigned risk, and conspicuous flame or thorn motifs for high assigned risk. The final column states the intended degree of caution. Generated appearances depend on the scene screenshot and assigned risk level.}
\end{table*}

Material character supplies an intended association, while visual prominence helps make the cue available during an activity. A high-risk texture can remain consistent with the current scene while standing out from its surroundings. The study examines perceived cue clarity, experience in VR, and activity around objects under this combined representation strategy.

\subsection{Implementation}
The system combines a ZED Mini Stereo Camera\footnote{\url{https://store.stereolabs.com/products/zed-mini}}, Unity\footnote{\url{https://unity.com/products/unity-engine}}, and a Meta Quest 3 headset\footnote{\url{https://www.meta.com/quest/quest-3/}}. All local processing and rendering ran on a workstation equipped with an Intel Xeon W-3275M processor\footnote{\url{https://www.intel.com/content/www/us/en/products/sku/193754/intel-xeon-w3275m-processor-38-5m-cache-2-50-ghz/specifications.html}}, an NVIDIA RTX 5080 graphics card\footnote{\url{https://www.nvidia.com/en-us/geforce/graphics-cards/50-series/rtx-5080/}}, and 192\,GB of memory. The headset was connected to the workstation via a wired connection. Red-green-blue (RGB) images and depth observations from the camera support live spatial reconstruction of the physical room. YOLO3D\footnote{\url{https://github.com/ruhyadi/yolo3d}} supplies object categories to a classification stage that assigns low, medium, or high risk based on material softness or hardness, sharp features, electrical and moving components, and potential heat generation (Section~\ref{sec:vocabulary}).

Texture generation begins after object recognition and risk classification. We used OpenAI's \texttt{gpt-image-2}\footnote{\url{https://developers.openai.com/api/docs/models/gpt-image-2}}, providing a screenshot of the virtual environment and the object's assigned risk level. The assigned level is an input to generation. The prompt explicitly asks the model to infer the scene's theme and combine material appearance, color, and contrast to convey that level within the scene. Generation thus implements the intended visual differentiation in C2. The prompt used for texture generation is provided in Appendix~\ref{app:texture-generation-prompt}.

Once mesh scanning is complete, Unity applies the generated texture directly to the reconstructed object, without manual selection among alternative outputs. The retextured mesh is rendered together with the virtual environment and displayed in the headset. The scanned mesh supplies shape and placement; the texture supplies apparent material and visual emphasis.

The three study conditions share live scanning, object recognition, and risk classification. Their rendering strategies determine whether a reconstructed surface retains its original appearance, receives a uniform environment texture, or receives Chameleon's risk-aware texture. This common pipeline supports the comparison of how physical objects are presented during the same virtual activity.

\section{Methods}\label{sec:study}
\begin{figure*}[!t]
\centering
\includegraphics[width=\textwidth]{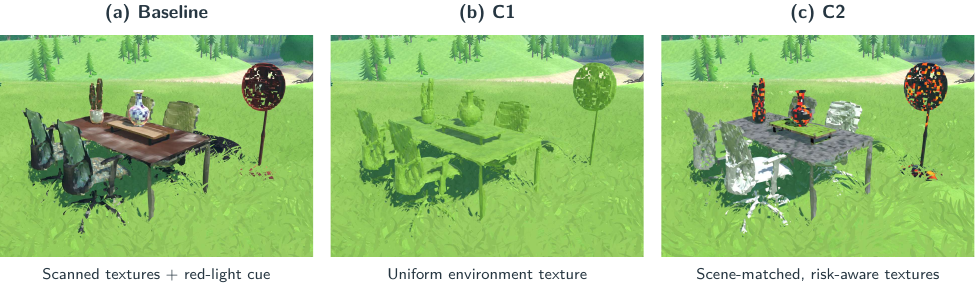}
\caption{System screenshots of the three rendering conditions: (a) Baseline preserves scanned object textures and adds a red-light cue to high-risk objects; (b) C1 applies a uniform environment texture; (c) C2 applies scene-matched textures differentiated by risk level. Each view shows the reconstructed table, chairs, tabletop objects, and standing fan in the woodland environment.}
\Description{Three side-by-side system screenshots show the same arrangement of reconstructed objects in a grassy woodland. Baseline shows a brown table, patterned chairs, distinct tabletop objects, and red emphasis on the fan. C1 renders the objects with a shared green environment texture. C2 shows a gray table, pale chairs, a green tabletop platform, and orange-and-black textures on the cactus, vase, and fan. Fragmented scanned mesh surfaces remain visible in all three views.}
\label{fig:conditions}
\end{figure*}

We conducted a three-condition, between-participants study to compare perceived risk communication, experience in VR, and activity around objects under three rendering strategies (Figure~\ref{fig:conditions}): scanned appearance with a high-risk red-light cue (Baseline), uniform environment-texture substitution (C1), and scene-matched, risk-aware textures (C2).

Baseline retains the physical objects' original textures captured by scanning, so users can draw on everyday knowledge of familiar objects and materials when judging contact risk. We selected this condition to compare risk-aware retexturing with judgments grounded in recognizable physical appearances. High-risk objects additionally receive red-light emphasis, following prior use of red overlays to identify hazardous kitchen-tool parts for people with low vision~\cite{Lee2024CookAR} and colored hazard zones or proximity cues to support awareness around collaborating robots~\cite{SanMartin2025HazardZones,Gruenefeld2020MindTheARm}.

C1 addresses integrating existing room contents into a themed virtual experience. Prior systems pair physical objects with contextually appropriate virtual counterparts~\cite{chameleon_B01}, match reconstructed surfaces to virtual lighting and color~\cite{chameleon_B08}, and transform furniture while preserving spatial and functional properties~\cite{chameleon_D01}. These approaches make physical surroundings part of the virtual setting while retaining a basis for interaction. We used uniform environment-texture substitution as a simplified reference for this integration goal, retaining scanned geometry and placement. Applying one texture across objects leaves material and risk differences unencoded and may reduce visual distinctiveness; we therefore assessed how this presentation relates to both scene experience and interaction around objects.

C2 applies scene-matched textures differentiated by assigned risk level, using apparent material, color, and contrast to convey the contact guidance defined in Section~\ref{sec:vocabulary}. Its comparison with C1 examines this combined visual differentiation relative to uniform scene matching, while Baseline provides a reference with recognizable appearances and explicit warnings. The comparisons evaluate each rendering strategy as a whole; texture generation method was not an experimental factor. Participants were randomly assigned to one of the three conditions.

Table~\ref{tab:conditions} compares the texture treatments used in each condition.

\begin{table*}[t]
\caption{Risk-aware retexturing and scene integration across the three rendering conditions.}
\label{tab:conditions}
\centering
\small
\renewcommand{\arraystretch}{1.2}
\begin{tabularx}{\textwidth}{@{}>{\raggedright\arraybackslash}p{0.34\textwidth}*{3}{>{\centering\arraybackslash}X}@{}}
\toprule
 & \textbf{Baseline} & \textbf{C1} & \textbf{C2} \\
\midrule
\textbf{Risk-aware retexturing} & No & No & Yes \\
\midrule
\textbf{Scene-matched retexturing} & No & Yes & Yes \\
\bottomrule
\end{tabularx}
\par\smallskip
\begin{minipage}{\textwidth}
\footnotesize
Baseline preserves original scanned textures to support risk judgments based on everyday knowledge of objects and materials, and adds a red-light warning for high-risk objects.
\end{minipage}
\Description{Two rows compare the texture treatments in Baseline, C1, and C2. Risk-aware retexturing is absent in Baseline and C1 and present in C2. Scene-matched retexturing is absent in Baseline and present in C1 and C2. A note explains that Baseline preserves original scanned textures to support risk judgments based on everyday knowledge of objects and materials, and adds a red-light warning for high-risk objects.}
\end{table*}

\subsection{Research Questions}
The study addressed three research questions:
\begin{itemize}
  \item \textbf{RQ1: Perceived risk communication.} How do participants rate their ability to distinguish risk levels and their sense of safety around objects under each representation?
  \item \textbf{RQ2: Experience in the scene.} How do reported presence, immersion, attention, and discomfort differ across representations during the task?
  \item \textbf{RQ3: Task performance and interaction.} How do estimated contacts with medium- and high-risk objects and closest approach to high-risk surfaces differ across representations?
\end{itemize}

\subsection{Participants}\label{sec:participants}
Our sample comprised 42 participants (14 per condition), recruited through campus email announcements and posters. Participants were aged 18--31 years ($M = 24.1$, $SD = 3.6$). Twenty-two participants identified as women and 19 as men; one did not report their gender. Twenty-nine participants had previously used a VR headset, while 13 had no prior VR experience. The study was approved by our university's Institutional Review Board.

\subsection{Procedure}
Participants first answered background questions about their VR experience, received task instructions, and completed headset fitting and tracking calibration. In their assigned condition, they then had ten minutes to touch as many of 200 virtual bubbles as possible, moving and reaching around the reconstructed objects. A researcher supervised the task, and participants could stop at any time. After removing the headset, participants completed the questionnaire and a 15-minute individual semi-structured interview.

\subsection{Measures and Data Collection}\label{sec:data-collection}
We assessed perceived risk communication, experience in VR, and activity around physical objects, with interviews providing accounts of cue interpretation and interaction decisions. Table~\ref{tab:measures} maps these measures to the research questions. Participants completed the 31 self-report items using sliders in Qualtrics\footnote{\url{https://www.qualtrics.com/}} after removing the headset. Study-specific items were treated as individual measures using available responses without imputation. The full wording of the study-specific items is provided in Appendix~\ref{app:questionnaire}.

\begin{table*}[t]
\caption{Study measures and their correspondence to the research questions.}
\label{tab:measures}
\centering
\small
\renewcommand{\arraystretch}{1.15}
\begin{tabularx}{\textwidth}{@{}>{\raggedright\arraybackslash}p{0.20\textwidth}>{\raggedright\arraybackslash}p{0.19\textwidth}>{\raggedright\arraybackslash}X>{\raggedright\arraybackslash}p{0.065\textwidth}@{}}
\toprule
Measure & Items or records & Content & RQ \\
\midrule
\multicolumn{4}{@{}l}{\textit{Perceived risk communication (RQ1)}} \\
Perceived safety & Q11.6--Q11.9 & Feeling safe, receiving sufficient obstacle cues, recognizing danger, and moving confidently & RQ1 \\
Perceived risk differentiation & Q11.10--Q11.13 & Judging relative danger, distinguishing risk levels, confidence in risk judgments, and misunderstanding danger & RQ1 \\
\midrule
\multicolumn{4}{@{}l}{\textit{Experience in the scene (RQ2)}} \\
Presence (IPQ) & Q27.1--Q27.13 & Spatial presence, involvement, and experienced realism & RQ2 \\
Immersive experience & Q11.1--Q11.5 & Feeling inside the virtual environment, believability, smoothness, task focus, and immersion & RQ2 \\
Physical discomfort & Q20.1--Q20.5 & Dizziness, nausea, eyestrain, sweating or increased heart rate, and disorientation or loss of balance & RQ2 \\
\midrule
\multicolumn{4}{@{}l}{\textit{Activity around objects (RQ3)}} \\
High-risk object contacts & Fingertip trajectories and object meshes & Number of estimated contact episodes with high-risk objects per session & RQ3 \\
Medium-risk object contacts & Fingertip trajectories and object meshes & Number of estimated contact episodes with medium-risk objects per session & RQ3 \\
Closest approach to high-risk objects & Fingertip trajectories and object meshes & Minimum distance from a tracked fingertip on either hand to a high-risk object surface during the task (cm) & RQ3 \\
\midrule
\multicolumn{4}{@{}l}{\textit{Qualitative accounts (RQ1--3)}} \\
Interpretations and experiences & Audio recordings & Interpretation of risk cues, engagement and distraction, and reasons for approaching, avoiding, or hesitating near objects & RQ1--3 \\
\bottomrule
\end{tabularx}
\Description{The table maps measures to three research questions: perceived safety and risk differentiation for RQ1; presence, immersive experience, and physical discomfort for RQ2; and estimated contacts with high- and medium-risk objects and closest approach to high-risk surfaces for RQ3. Interview accounts address all three questions. Columns identify each measure, its questionnaire items or records, its content, and the research question.}
\end{table*}

\subsubsection{Perceived Risk Communication}
For RQ1, eight study-specific items assessed participants' perceived ability to interpret risk cues and their sense of safety around physical objects. Four addressed feeling safe, the sufficiency of obstacle cues, danger recognition, and movement confidence. Four addressed relative danger, differentiation of risk levels, confidence in risk judgments, and reported misunderstanding of danger. Safety ratings address reassurance when acting around physical obstacles, a concern highlighted by research on collision anxiety in VR~\cite{chameleon_A22}. Risk-differentiation ratings address the interpretation of graded hazard information, as examined in hazard-awareness displays~\cite{SanMartin2025HazardZones}.

\subsubsection{Presence and Immersive Experience}
We assessed presence (RQ2) with the Igroup Presence Questionnaire (IPQ)~\cite{Schubert2001Presence}. The 13 items covered spatial presence (five), involvement (four), and experienced realism (four). Each subscale score was the mean of its answered items after reverse coding~\cite{IPQScoring}, requiring at least half of its items to be answered. Five study-specific items further assessed feeling inside the virtual environment (VE), believability, presentation and interaction smoothness, task focus, and immersion. Work on integrating physical distractions and guiding attention in VR examines how cues can fit an experience while limiting disruption~\cite{Tao2022Integrating,Lange2020HiveFive}. Our study-specific items address this concern through perceived coherence, interaction fluency, and task engagement.

\subsubsection{Physical Discomfort}
Five items assessed discomfort during the task (RQ2): dizziness, nausea, eyestrain, sweating or increased heart rate, and disorientation or loss of balance. Dizziness, nausea, eyestrain, and sweating are also assessed by the Simulator Sickness Questionnaire~\cite{Kennedy1993SSQ}. Background questions also covered previous VR use, movement during VR, and prior VR-related nausea or balance difficulties.

\subsubsection{Contact and Approach by Object Risk}
For RQ3, we characterized interaction around objects through estimated contact episodes and closest approach, using fingertip trajectories registered to reconstructed meshes. A contact episode required a fingertip to remain within 5\,cm of a sampled surface for at least 0.2 seconds, with gaps of up to 0.3 seconds merged. We summed episodes across available objects within each session, separately for high- and medium-risk categories, and measured closest approach as the minimum distance in centimeters from a fingertip on either hand to any sampled high-risk surface during the task. 

To place these risk-specific measures in the context of overall activity, we also recorded task progress and typical proximity to objects. Task progress was the number of successful bubble touches within the ten-minute window delimited by countdown-start and countdown-end events. Typical head and hand proximity was summarized separately using the time-weighted median Euclidean distance to the nearest object transform origin. These summaries used positions sampled at approximately 0.1-second intervals and required at least 80\% valid distance coverage.

\subsubsection{Semi-Structured Interviews}
After the questionnaire, we conducted an individual semi-structured interview lasting approximately 15 minutes. The interviews explored which visual cues participants relied on and why they approached, avoided, or hesitated near particular objects. We asked participants to recall moments when they felt safe or needed extra caution, explain how they judged an object's risk, and describe uncertainty or changes of mind. Questions about immersion addressed moments of engagement, distraction, and whether the presentation felt natural. Follow-up questions concerned the participant's assigned condition: associations with scanned appearance in Baseline, interpreting objects that shared one texture in C1, and the meaning and scene fit of risk-differentiated textures in C2. Interviews were audio recorded to preserve participants' explanations for subsequent qualitative analysis.

\subsection{Data Analysis}\label{sec:analysis}
We conducted exploratory analyses with retrospectively specified scoring and comparison families. We assessed IPQ internal consistency using complete-case Cronbach's $\alpha$ and checked subscale distributions within each condition using Shapiro--Wilk tests. Spatial presence departed from normality in C2. We used Kruskal--Wallis tests for overall condition differences and two-sided Mann--Whitney $U$ tests for all three pairwise comparisons. Individual Likert-type items and behavioral measures were also compared using these nonparametric tests.

We applied Benjamini--Hochberg (BH) correction~\cite{BenjaminiHochberg1995FDR} using five outcome domains: three IPQ subscales, five experience items, eight safety/risk items, five symptom items, and three task/proximity measures. Omnibus $p$-values were corrected jointly within each domain. For pairwise tests, each domain and condition pair defined a family: for example, the eight safety/risk tests comparing C2 with C1 were corrected together. The four outcomes in Figure~\ref{fig:risk-behaviour} formed an additional family for each condition pair. Adjusted values are denoted by $q$, with $q<.05$ indicating significance. We report Cohen's $d$.

Pointwise 95\% confidence intervals (CIs) for means and mean differences used 20,000 bootstrap resamples. To assess sensitivity to the comparison method, we also used Welch tests and 10,000 permutations of mean differences~\cite{Winkler2014Permutation}, applying the same families of outcomes for each condition pair as in the primary tests. Further checks used global BH and Holm corrections~\cite{Holm1979Sequential} across the 72 questionnaire and task/proximity comparisons, complete IPQ responses, and questionnaire models adjusted for motion-discomfort history with heteroskedasticity-robust standard errors using the HC3 estimator. These checks assessed the dependence of findings on multiplicity, missing responses, and participant background.

Three researchers participated in the qualitative analysis. We manually coded transcript excerpts concerning risk judgments, physical confirmation, and task engagement, retaining participant identifiers and timestamps to connect each code to its transcript context. Related codes were grouped into broader categories and themes reflecting our interpretation of how participants used visual cues, formed expectations about physical properties, and decided how to interact with objects. We jointly reviewed the coding and themes, referring disagreements to the third researcher for a final decision. We compared accounts within and across rendering conditions, considering shared patterns and contrasting interpretations, and organized the findings by condition. Questionnaire and behavioral patterns provided context for interpreting participants' explanations.

\section{Results}\label{sec:results}
We present quantitative comparisons of reported perceptions and reconstructed activity, followed by interview findings on how participants interpreted the cues and judged physical actions.

\subsection{Quantitative Results}\label{sec:quantitative-results}
Our Kruskal--Wallis tests identified differences in safety and risk-cue ratings, task experience, and discomfort. We report pairwise comparisons and risk-specific contact and approach estimates below; Table~\ref{tab:results} in Appendix~\ref{app:quantitative-results} provides the descriptive statistics and adjusted comparisons.

\begin{figure*}[!tp]
\centering
\includegraphics[width=0.9\textwidth]{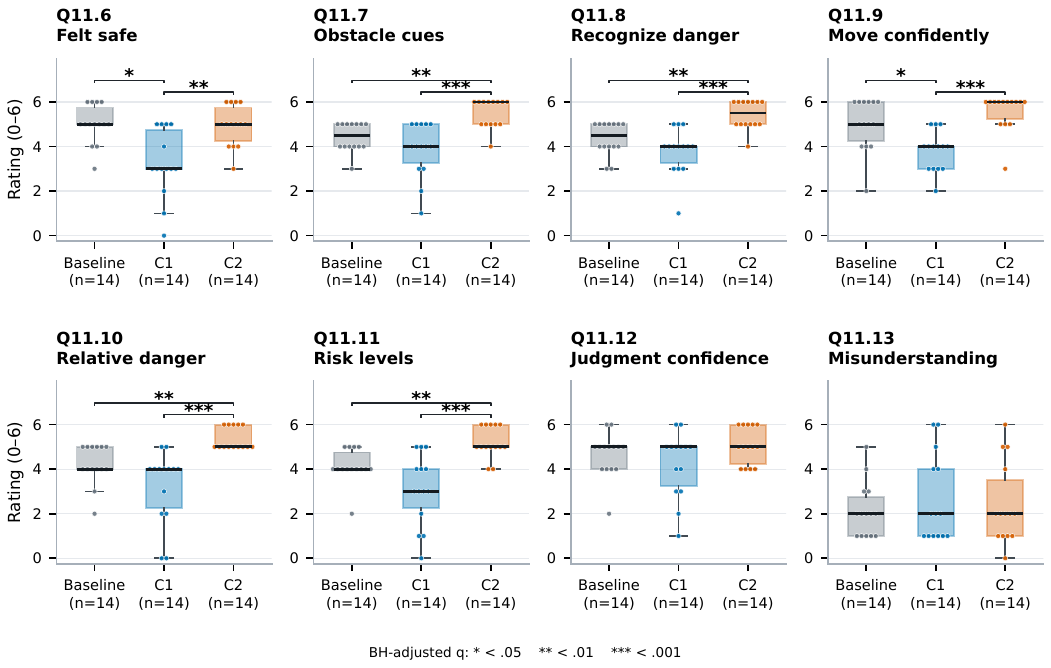}
\caption{Individual safety and risk ratings ($n=14$ per condition). Boxes span the interquartile range (IQR), lines mark medians, and whiskers extend to the most extreme observations within 1.5 IQR of the box. Dots show individual responses. Higher scores indicate more favorable perceptions except for misunderstanding (Q11.13). Brackets indicate two-sided Mann--Whitney $U$ tests with BH adjustment across eight items for each condition pair: $^{*}q<.05$, $^{**}q<.01$, $^{***}q<.001$.}
\Description{Eight panels compare safety and risk ratings across Baseline, C1, and C2 using boxplots and individual response dots. Boxes show the middle half of responses and their medians; whiskers extend to observations within 1.5 interquartile ranges. Dots include observations beyond the whiskers. C2 has higher ratings than C1 on the first six items and Baseline on four cue-interpretation items. Judgment confidence and misunderstanding have no adjusted significant comparison; higher misunderstanding scores indicate more reported misunderstanding.}
\label{fig:safety-items}
\end{figure*}

\subsubsection{Higher Ratings for Perceived Cue Clarity}
Participants in C2 reported clearer risk cues than those in either comparison condition (Figure~\ref{fig:safety-items}). This pattern covered sufficient obstacle cues, recognition of potential danger, understanding relative danger, and distinguishing risk levels (Q11.7, Q11.8, Q11.10, and Q11.11). All four items favored C2 over C1 (all $q<.001$) and Baseline (all $q<.003$), with agreement from the adjusted Welch and mean-permutation tests. For example, danger-recognition ratings averaged 5.43 in C2, 3.79 in C1, and 4.36 in Baseline. The C2--C1 difference was 1.64 points on the 0--6 scale (95\% CI [1.07, 2.29], $d=1.88$). All eight C2 comparisons remained significant after adjustment for motion-discomfort history.

None of the four cue-interpretation items differed significantly between C1 and Baseline (all $q>.12$).

Differences in judgment confidence and reported misunderstanding were inconclusive. Judgment confidence (Q11.12) averaged 5.07 in C2, 4.21 in C1, and 4.64 in Baseline, with no significant pairwise comparison (all $q>.14$). Agreement with sometimes misunderstanding an obstacle's danger (Q11.13) averaged 2.43, 2.64, and 2.14, respectively (all $q>.79$); higher values indicate more misunderstanding. Neither item differed significantly after covariate adjustment.

\subsubsection{Perceived Safety and Movement Confidence}
Uniform substitution received lower ratings for feeling safe and moving confidently than either alternative. Perceived safety (Q11.6) averaged 3.21 in C1, compared with 4.93 in C2 and 5.00 in Baseline. The C2--C1 comparison ($q=.003$, $d=1.36$) and Baseline--C1 comparison ($q=.012$) were significant, whereas C2--Baseline was inconclusive ($q=.905$). Movement confidence (Q11.9) showed a similar pattern: C2 ($M=5.57$) and Baseline ($M=5.00$) exceeded C1 ($M=3.79$; $q<.001$ and $q=.018$, respectively), with no significant C2--Baseline difference ($q=.187$).

All four comparisons with C1 remained significant in the supporting tests and after covariate adjustment. Global Holm correction yielded $p=.141$ for C2--C1 perceived safety. The ratings distinguish perceptions of cue clarity from feeling safe and moving confidently around the represented objects.

\begin{figure*}[!tp]
\centering
\includegraphics[width=0.9\textwidth]{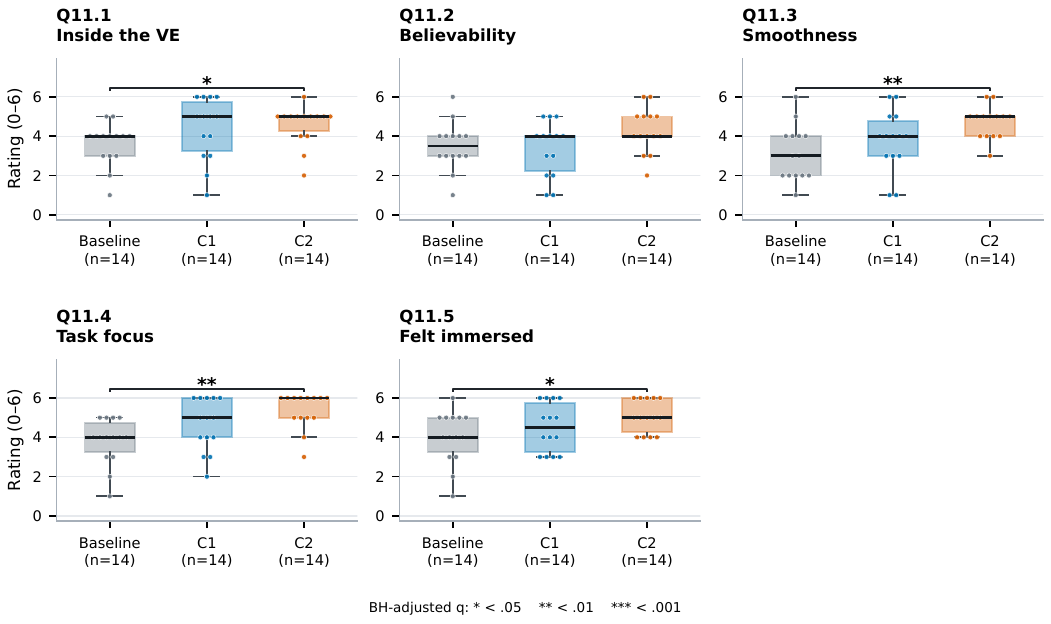}
\caption{Individual experience ratings ($n=14$ per condition), reported separately from IPQ. Boxes show medians and IQRs, whiskers extend to observations within 1.5 IQR, and dots show individual responses. Brackets indicate two-sided Mann--Whitney $U$ tests with BH adjustment across five items per condition pair: $^{*}q<.05$, $^{**}q<.01$, $^{***}q<.001$.}
\Description{Five panels compare feeling inside the environment, believability, smoothness, task focus, and feeling immersed using boxplots and individual response dots. C2 has higher ratings than Baseline on feeling inside, smoothness, task focus, and feeling immersed in the primary adjusted tests. Believability and all comparisons involving C1 are nonsignificant. Dots include observations beyond the whiskers.}
\label{fig:experience-items}
\end{figure*}

\subsubsection{Scene Experience and Presence Ratings}
Participants in C2 reported stronger feelings of being inside the environment and feeling immersed than those in Baseline (Figure~\ref{fig:experience-items}). Feeling inside the environment (Q11.1) averaged 4.57 in C2 versus 3.57 in Baseline ($q=.014$, $d=0.95$). Feeling immersed (Q11.5) averaged 5.07 versus 4.00 ($q=.037$, $d=0.95$), with C1 averaging 4.50. Both C2--Baseline differences remained significant in the adjusted Welch and mean-permutation tests and after covariate adjustment. Global BH correction across all 72 comparisons yielded $q=.088$ for direct immersion. Neither C2--C1 nor C1--Baseline was significant on either item. Believability (Q11.2) did not differ significantly for any pair (all $q>.08$), although C2--Baseline became significant after covariate adjustment ($q_{HC3}=.042$).

Complete-case Cronbach's $\alpha$ was .82 for spatial presence, .80 for involvement, and .84 for realism.

\begin{figure*}[!tp]
\centering
\includegraphics[width=0.9\textwidth]{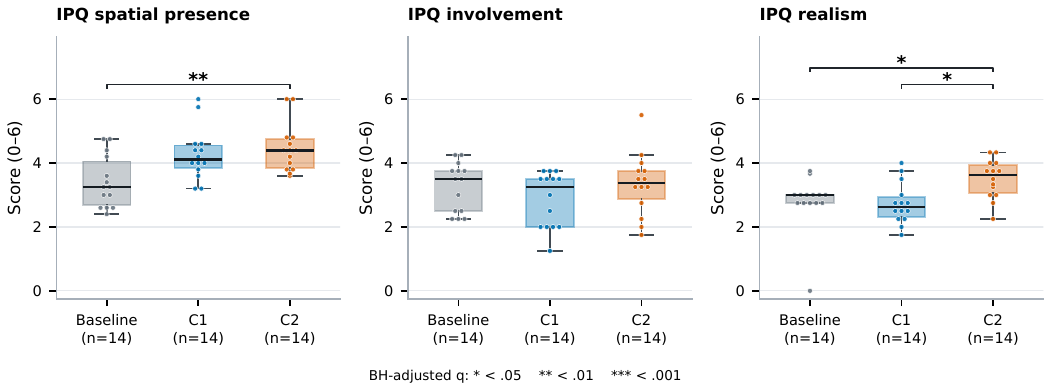}
\caption{IPQ subscale scores ($n=14$ per condition). Boxes show medians and IQRs, whiskers extend to observations within 1.5 IQR, and dots show individual scores. Scores require at least half of a subscale's items to be answered. Brackets indicate two-sided Mann--Whitney $U$ tests with BH adjustment across three subscales per condition pair: $^{*}q<.05$, $^{**}q<.01$, $^{***}q<.001$.}
\Description{Three panels compare spatial-presence, involvement, and realism scores using boxplots and individual response dots. C2 has higher spatial-presence scores than Baseline and higher realism scores than both comparison conditions. No involvement comparison is significant after adjustment. Dots include observations beyond the whiskers.}
\label{fig:ipq-results}
\end{figure*}

Figure~\ref{fig:ipq-results} shows the IPQ subscale comparisons. Spatial-presence scores were higher in C2 ($M=4.47$, $SD=0.84$) than in Baseline ($M=3.29$, $SD=0.82$; $q=.003$, $d=1.42$). C1 averaged 3.98, with no significant difference from either C2 ($q=.201$) or Baseline ($q=.252$). Realism scores were higher in C2 ($M=4.07$) than in C1 ($M=2.98$; $q=.023$, $d=1.18$) or Baseline ($M=2.80$; $q=.004$, $d=1.33$). All three significant IPQ comparisons passed the supporting tests. C1 and Baseline did not differ significantly in realism ($q=.754$), and no involvement comparison was significant (all $q>.20$).

All three significant IPQ comparisons remained significant after covariate adjustment (all $q_{HC3}<.020$) and when requiring complete subscales (all $q<.027$).

\subsubsection{Interaction Fluency and Task Focus}
C2 received higher ratings than Baseline for the experience of carrying out the task. Participants rated presentation and interaction as smoother and more natural (Q11.3) in C2 than in Baseline (4.71 versus 3.14; $q=.007$, $d=1.36$). They also reported greater task focus without interface distraction (Q11.4; 5.36 versus 3.79; $q=.003$, $d=1.47$). Both differences remained significant in the supporting tests and after covariate adjustment; task focus also survived global Holm correction ($p=.045$). C1's means were intermediate (3.79 for smoothness and 4.64 for focus), with no significant difference from either condition on either item. The reported advantage thus concerned C2 relative to scanned appearance with a high-risk warning.

\begin{figure*}[!tp]
\centering
\includegraphics[width=0.9\textwidth]{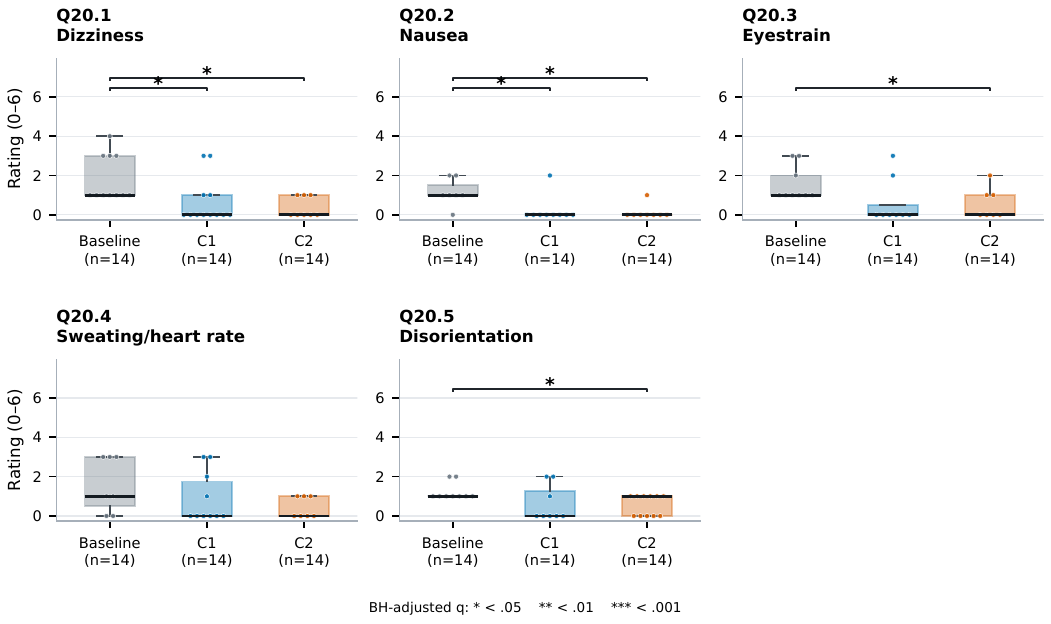}
\caption{Individual symptom ratings ($n=14$ per condition); higher values indicate more discomfort. Boxes show medians and IQRs, whiskers extend to observations within 1.5 IQR, and dots show individual responses. Brackets indicate two-sided Mann--Whitney $U$ tests with BH adjustment across five items per condition pair: $^{*}q<.05$, $^{**}q<.01$, $^{***}q<.001$.}
\Description{Five panels compare dizziness, nausea, eyestrain, sweating or increased heart rate, and disorientation using boxplots and individual response dots. Four C2--Baseline and two C1--Baseline primary comparisons are significant; no C2--C1 comparison is significant. Dots include observations beyond the whiskers.}
\label{fig:symptom-items}
\end{figure*}

\subsubsection{Physical Discomfort}
Lower symptom ratings were concentrated in the C2--Baseline comparison (Figure~\ref{fig:symptom-items}). Dizziness, nausea, and disorientation were lower in C2 ($q=.014$, .018, and .019, respectively), with agreement across the primary and supporting tests. For example, dizziness averaged 0.38 in C2 and 1.82 in Baseline ($d=-1.51$). Eyestrain also favored C2 in the primary test ($q=.043$), but not the adjusted permutation test ($q_P=.076$). Sweating or increased heart rate did not differ significantly. C1 received lower dizziness and nausea ratings than Baseline (both $q=.033$), but neither comparison passed the supporting tests; the other C1--Baseline comparisons and all C2--C1 comparisons were inconclusive.

Symptom models used available responses and included history of motion discomfort as a covariate. With HC3 standard errors, all pairwise comparisons yielded $q\geq.05$.

\subsubsection{Greater Estimated Clearance around High-Risk Objects}
\begin{figure*}[!tp]
\centering
\includegraphics[width=\textwidth]{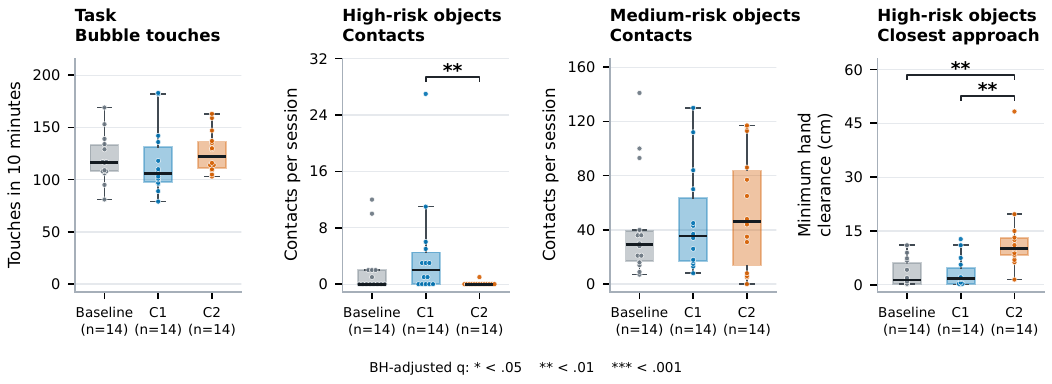}
\caption{Task progress and estimated contact and approach ($n=14$ per condition), using fingertip trajectories relative to sampled mesh surfaces. Boxes show medians and IQRs, whiskers extend to observations within 1.5 IQR, and dots show individual sessions. Brackets indicate two-sided Mann--Whitney $U$ tests with BH adjustment across four outcomes per condition pair: $^{*}q<.05$, $^{**}q<.01$, $^{***}q<.001$.}
\Description{Four panels show bubble touches, estimated high-risk and medium-risk contact episodes, and minimum distance to high-risk surfaces using boxplots and individual session dots. C2 has fewer estimated high-risk contacts than C1 and greater minimum high-risk distance than both alternatives. Medium-risk contact estimates and bubble touches do not differ significantly. Dots include observations beyond the whiskers.}
\label{fig:risk-behaviour}
\end{figure*}

Contact estimates and closest approach addressed activity around objects by risk category (RQ3; Figure~\ref{fig:risk-behaviour}). Estimated high-risk contact episodes averaged 0.07 in C2, 4.29 in C1, and 2.07 in Baseline. C2 had fewer estimated episodes than C1 ($q=.003$); its comparison with Baseline did not reach significance ($q=.0505$). Medium-risk contact estimates averaged 53.64, 46.57, and 42.29, respectively, with no significant pairwise differences (all $q>.56$). Minimum distance to high-risk surfaces was greater in C2 (12.85\,cm) than in C1 (3.29\,cm; $q=.003$) or Baseline (3.17\,cm; $q=.002$).

Together, these estimates associate C2 with fewer high-risk contact episodes than C1 and greater high-risk surface clearance than both alternatives.

\subsubsection{Task Progress and Object Proximity}
Supplementary measures characterized overall activity during the task. In the task-progress panel of Figure~\ref{fig:risk-behaviour}, participants touched an average of 126.50 bubbles in C2 ($SD=19.66$), 117.43 in C1 ($SD=32.23$), and 120.07 in Baseline ($SD=23.06$) during the ten-minute task. No comparison was significant (C2--C1: $q=.137$; C2--Baseline: $q=.566$; C1--Baseline: $q=.883$).

Mean session-level head proximity was 1.217 scene units in C2, 1.203 in C1, and 1.215 in Baseline; corresponding tracked-hand values were 0.799, 0.765, and 0.814. No comparison was significant (all $q>.34$). These time-weighted session medians summarize typical distance to object reference points, complementing the risk-specific measures of surface contact and closest approach.

\FloatBarrier
\subsection{Qualitative Results}\label{sec:qualitative-results}
Our interview analysis highlighted how physical contact helped establish confidence in familiar objects, visual coherence coexisted with uncertain object properties, and salient risk cues left questions about how to act.

\subsubsection{Baseline: Familiar Objects and Uncertain Boundaries}\label{sec:qual-baseline}
Recognizable objects provided an initial basis for judging risk. Everyday knowledge distinguished familiar tables and chairs from potentially harmful cacti, alongside the red warning applied to particular objects (P41). Convincing object appearance did not always extend to the scene as a whole: office furniture could look realistic while seeming out of place on grass (P34).

Confidence in a familiar object's physical presence still depended on contact: ``\textit{I didn't trust the table until I touched it and I felt a hard table}'' (P42). Contact could also change the perceived risk category. Objects already touched were considered low risk, whereas ``\textit{If I hadn't touched them yet, I would also think of them as medium risk}'' (P32). These judgments reflected whether an encounter had been physically confirmed, alongside expectations about the object itself.

Uncertainty persisted where physical boundaries were difficult to anticipate. The apparently open outdoor scene obscured the limits of the room until walls became apparent on approach (P32). Task demands also prompted changes in how familiar surfaces were used. Failed attempts to reach a bubble from the floor led to climbing onto the table, with kneeling chosen over standing for greater perceived stability (P36). Familiar appearance thus supported initial expectations, while contact, boundary awareness, and posture shaped confidence in movement.

\subsubsection{C1: Visual Coherence with Uncertain Object Properties}\label{sec:qual-uniform}
A shared texture could make the simplified scene feel coherent: ``\textit{Because everything had the same visual style, it felt quite harmonious}'' (P14). This consistency coexisted with uncertainty about object identity and material properties.

Uncertainty about what an object was could discourage contact: ``\textit{When I don't know what an object is, I won't touch it}'' (P15). A recognizable table, by comparison, could feel safe despite the possibility of collision because its identity was understood (P15). Recognition still left some physical properties unresolved. Knowing that a fan or vase might tip over did not make fragility apparent: ``\textit{I didn't notice it was glass}'' (P24). Similarly, a toaster was identified through touch, with its cold metal surface conflicting with the common texture displayed across objects (P26).

Task focus could reduce attention to these visual details: ``\textit{I did not mind the surroundings much. I was just finding the bubble}'' (P23). Movement still involved selective strategies, such as hopping over the beanbag while walking around the fan (P26). The setting supplied additional reassurance through expectations that researchers would exclude objects capable of serious harm (P16) and the belief that greater caution would be needed outside the supervised setting (P19). Willingness to interact therefore drew on physical confirmation and contextual reassurance alongside visual interpretation.

\subsubsection{C2: Salient Risk Cues and Action-Specific Confidence}\label{sec:qual-risk}
Color associations provided an immediate basis for caution: ``\textit{I just associated red with bad}'' (P07). Red objects were easy to notice and avoid (P04), while an orange-and-black fan appeared more dangerous than the gray table and armchairs (P01). Differentiated appearance could therefore indicate where extra care was needed.

Visual differences did not always convey an interpretable warning: ``\textit{I don't know the meaning from the textures}'' (P10). A red plant conflicted with expectations that plants would be green. Suggested clarifications included object names, risk levels, and possible harms (P10). Uncertainty also concerned what ``dangerous'' meant in practice, such as whether an object was sharp or could cause injury on contact (P12). Detecting a cue did not necessarily explain which property or action made the object hazardous.

Confidence remained specific to the intended action. Confirming the table's size and position supported tracing its edge, but uncertainty about stability discouraged leaning on it (P12). High confidence in judging danger could also coexist with untested assumptions: ``\textit{I don't know if it's true because I didn't touch anything looks dangerous}'' (P11). Avoidance allowed action on a warning without resolving whether the judgment was accurate.

Scene fit depended on how the objects and setting were interpreted together. Stone tables and counters could suggest a magical forest inhabited by someone (P07), yet a recognizable piece of furniture could still seem incongruous: ``\textit{a very modern table just in the middle of the forest}'' (P13). Matching surface appearance did not always explain why an object belonged there.

Task motivation and supervision also shaped perceived safety. Pursuing a higher score could draw attention away from physical walls (P02), while feeling protected could rest on the assurance that ``\textit{the interviewer said he will keep me safe}'' (P11). Visible warnings therefore operated alongside the activity and its perceived safeguards.

\section{Discussion}\label{sec:discussion}
The findings distinguish noticing a caution cue, interpreting a physical property, and judging an intended action. We develop these distinctions from the interview accounts alongside the perceived-clarity comparisons, then consider their implications for evaluation and design.

\subsection{Principal Results}
\subsubsection{Visual Cues Guide Initial Caution}
C2 received higher ratings than both alternatives for four aspects of perceived cue clarity (RQ1). Prior work on affordance augmentation found that participants with low vision preferred cues distinguishing grabbable and hazardous tool parts to whole-object overlays~\cite{Lee2024CookAR}. Our interviews identify a related need for information about contact: familiar warning colors prompted caution, while participants sought explanations of the physical concern, such as sharpness or another source of harm (Section~\ref{sec:qual-risk}). These accounts distinguish noticing where care is appropriate from understanding why. Together, the findings suggest that differentiated cues should communicate both the location of a concern and the physical property relevant to the intended action.

The reconstructed activity associated C2 with greater clearance from surfaces assigned high risk than both alternatives and fewer estimated high-risk contact episodes than C1 (RQ3). Medium-risk contacts, bubble touches, and typical head and hand proximity showed no significant pairwise differences. The significant behavioral comparisons therefore concern contact estimates and closest approach within the high-risk category. This pattern is consistent with participants' descriptions of noticing and avoiding conspicuous objects (Section~\ref{sec:qual-risk}).

\subsubsection{Confidence Depends on the Intended Use}
Physical confirmation shaped participants' judgments about how they could use a represented surface. In Baseline, touching the table confirmed that it was physically present. In C2, confirming a table's size and position supported tracing its edge while uncertainty about stability discouraged leaning on it (Section~\ref{sec:qualitative-results}).

Substitutional-reality experiments found that material-related mismatches involving touch, temperature, and weight affected the perceived compatibility of physical and virtual objects~\cite{chameleon_B01}. Our accounts clarify how such expectations enter decisions about particular uses: spatial confirmation can support tracing an edge, while leaning requires confidence in stability and support. This connects prior findings about acceptable substitution to the properties users need to verify before acting on a represented surface.

\subsubsection{Scene Fit and Material Meaning Shape Interpretation}
C2 received higher smoothness, task-focus, and self-reported immersion ratings than Baseline (RQ2), with no significant differences from C1 on these items. IPQ spatial-presence scores were higher in C2 than Baseline, and experienced-realism scores were higher than in both alternatives. Prior work found that incorporating real-world distractions into a virtual scene improved presence compared with using noise-canceling headphones~\cite{Tao2022Integrating}. Themed reconstructions in Roomify likewise received higher presence ratings than passthrough and fully virtual comparison conditions~\cite{chameleon_D01}. Our experience findings follow a similar direction, with scene-integrated appearances receiving favorable ratings relative to scanned objects with an explicit warning.

The interviews add a distinction between visual coherence and physical meaning. A common texture could make the scene feel harmonious while concealing differences between glass and metal. C2's differentiated surfaces could suggest a magical forest, yet recognizable office furniture could still seem out of place (Sections~\ref{sec:qual-uniform} and~\ref{sec:qual-risk}). These accounts qualify what scene matching involves: its interpretation depends on the material properties and role suggested by the object as well as its agreement with the surrounding palette.

\subsection{Methodological Reflections}
\subsubsection{Comparing Complete Rendering Strategies}
The conditions compare three ways to present reconstructed objects. Baseline retains original scanned textures and adds an explicit warning, C1 applies a common environment texture, and C2 deliberately combines scene-related material associations, color, and contrast to differentiate assigned risk categories. The contribution is empirical evidence about how users experience and interpret these complete presentations. The observed differences apply to the combined C2 appearance; they do not identify an independent effect of material meaning, visual prominence, or image generation. C1 situates this presentation relative to uniform scene matching, while Baseline provides a reference for risk judgments grounded in everyday knowledge of familiar objects and materials, supplemented by an explicit warning.

\subsubsection{Evaluating Risk at the Level of an Encounter}
The assigned categories organize the display and the analysis of activity around objects. Questionnaire ratings describe perceived clarity and confidence, while geometry-based measures describe fingertip proximity to reconstructed surfaces. Together with interview accounts, these observations connect the intended cue to users' appraisals and interaction patterns. We interpret the behavioral findings as patterns of approach and estimated contact during the supervised task. Establishing that the categories correspond to physical hazards requires an assessment of the objects and intended actions independent of their textures.

An encounter provides a useful unit for organizing these observations: the displayed cue refers to an object with particular physical properties, and the user considers that object while pursuing an action. Relating these elements makes it possible to examine how a cue is used when passing, touching, or leaning on a surface. This perspective connects risk communication to the physical demands of the activity.

\subsubsection{Interpreting Safety in a Supervised Task}
The bubble task made reaching and movement observable while participants pursued a virtual goal. It also made task motivation and supervision part of the experience being measured. Interviews connected reduced attention to surroundings with score seeking and attributed reassurance to the researcher (Section~\ref{sec:qualitative-results}). Related work likewise shows that users may deliberately cross VR boundaries using their knowledge of the physical surroundings~\cite{chameleon_A05}. A response to a warning consequently needs to be understood in relation to the user's goal and expectations about the room~\cite{Kim2025On}.

Evaluations should document the safety briefing and distinguish confidence in the representation from confidence in the study's safeguards. Repeated sessions with different task demands and consistent supervision could examine whether the same cues remain useful as users become familiar with the environment.

\Needspace{8\baselineskip}
\subsection{Implications for Designing Risk-Aware Substitution}
\subsubsection{Relate Caution to the Intended Action}
Our interview accounts show how physical expectations enter judgments about particular uses. We suggest specifying the action a cue is intended to inform before selecting its visual material. For a table, a cue about incidental contact would need to communicate different properties from a cue about using the surface for support. Object-level categories can provide an initial warning while leaving this action-specific information to be resolved.

Affordance augmentations provide a relevant precedent by distinguishing grabbable and hazardous parts of kitchen tools~\cite{Lee2024CookAR}. For risk-aware retexturing, the surface cue could similarly refer to the action under consideration~\cite{Xu2024SafeRDW}. Incidental contact, grasping, and supporting body weight involve different properties, even on the same object~\cite{Lin2020ARchitect,chameleon_B09,Jin2026BehaviorAware}. A representation could identify a region suitable for the intended contact and indicate when the suitability of a use is unknown. Candidate cues should be evaluated through the actions users expect them to permit~\cite{Murillo2017ErgO,Lu2026MoveTogether}, alongside recognition of the assigned category.

\subsubsection{Preserve Physical Meaning within a Coherent Scene}
The interviews suggest that scene fit involves more than a shared palette. Uniform textures could make objects look harmonious while concealing material distinctions, and a retextured office table could still seem out of place in a forest (Sections~\ref{sec:qual-uniform} and~\ref{sec:qual-risk}). These accounts bring the substitutional-reality question of acceptable mismatch~\cite{chameleon_B01} into the interpretation of risk: a visually plausible substitution can also alter expectations about physical contact.

Prior work integrates real-world stimuli by giving them an explanation within the virtual scene~\cite{Tao2022Integrating}, while the Haptic Fidelity Framework treats agreement between touch and other sensory information as a component of fidelity~\cite{chameleon_B11}. For Chameleon, these perspectives suggest selecting textures for both their place in the scene and the material properties they imply. A stone-like surface may fit a woodland setting while implying stability that a movable object does not possess. Designers should therefore assess expected hardness, mobility, and contact use when screening generated appearances. Comparisons that hold visual prominence constant while varying material meaning would help determine which aspects support interpretation and which introduce misleading expectations.

\subsubsection{Support Verification as Objects and Tasks Change}
Existing systems offer complementary ways to support inspection: Roomify enables correction of generated placements and requires confirmation of flagged objects before entry into VR~\cite{chameleon_D01}, while GradualReality adapts the amount of physical-world information shown to the user's interaction context~\cite{Seo2024GradualReality}. Building on these approaches, risk-aware substitution could let users inspect the physical object~\cite{chameleon_C07} or request a clearer explanation when preparing to interact. Detected movement or uncertain classification could trigger renewed verification before the appearance is again treated as dependable. Such changes should target the object and moment requiring attention so that verification remains usable during the task.

\subsection{Limitations and Future Work}\label{sec:limitations}
Our evaluation covered a single woodland vocabulary in a short, supervised reaching task. Roomify has demonstrated style transformations that preserve key geometric and functional properties across themed environments~\cite{chameleon_D01}; our study examined how risk-differentiated surface appearances were interpreted within one such setting. A next step is to evaluate generated cues across scene themes and repeated sessions, testing whether material associations remain interpretable as the setting and users' familiarity change. Repeating generation for the same object and risk category would also help assess consistency across texture outputs.

The rendering conditions jointly varied material associations, visual variety, and prominence. Earlier substitutional-reality experiments manipulated material and geometric mismatch~\cite{chameleon_B01}, providing a basis for more controlled comparisons of these properties. A factorial study could vary material meaning and visual prominence independently, including conspicuous cues without risk-specific material associations, while keeping object geometry and texture variety comparable. Such comparisons would help distinguish responses to generic visual differentiation from responses to the material properties depicted. The analyses were exploratory; future comparisons should specify outcomes and correction families in advance.

Independent assessment of the classification should document each object's physical properties and operating state, specify the intended contact action, and compare the assigned category with judgments made without viewing its generated texture. This would separate agreement about the physical concern from interpretation of the visual cue. Existing systems also provide concrete approaches for extending verification during interaction: Roomify requires confirmation of flagged placements before VR entry~\cite{chameleon_D01}, and GradualReality changes representations as users approach and manipulate objects~\cite{Seo2024GradualReality}. Future work could combine textures with inspection on request and cues for particular object regions, comparing users' action judgments with independently assessed contact properties. Synchronized recordings of displayed cues, mesh updates, fingertip trajectories, and observed contacts would support evaluating how movement and update timing affect these judgments and the agreement between reconstructed contact estimates and observed interaction.

\Needspace{10\baselineskip}
\section{Conclusion}
We examined how users perceived and interpreted Chameleon's scene-matched risk cues during a reaching task. C2 received higher ratings for perceived cue clarity than both comparison conditions. Interview accounts linked visible warnings to initial caution and physical confirmation to judgments about particular uses. Participants distinguished locating a surface from judging its suitability for leaning, connecting confidence to the properties relevant to an intended action. These findings motivate risk cues that explain contact-relevant properties and support verification during interaction. The study contributes a comparison of perceived cues and an account of how visual appearance, physical confirmation, and task context inform interaction judgments.

\bibliographystyle{ACM-Reference-Format}
\bibliography{references}
\appendix
\section{Detailed Quantitative Results}\label{app:quantitative-results}
\begin{table*}[t]\centering\footnotesize\setlength{\tabcolsep}{4pt}
\caption{Questionnaire and task/proximity outcomes ($N=42$; $n=14$ per condition). Values are means (SD) and BH-adjusted two-sided Mann--Whitney $q$ values; bold marks $q<.05$. Bubble-touch comparisons use the four-outcome family in Figure~\ref{fig:risk-behaviour}; other families are defined in Section~\ref{sec:analysis}. Higher Q11.13 indicates more misunderstanding, and higher Q20 indicates more symptoms.}
\label{tab:results}\begin{tabular}{@{}lrrrrrr@{}}\toprule
& \multicolumn{3}{c}{Mean (SD)} & \multicolumn{3}{c}{Adjusted pairwise $q$} \\
Outcome & Baseline ($n=14$) & C1 ($n=14$) & C2 ($n=14$) & C2--C1 & C2--Base. & C1--Base. \\\midrule
\multicolumn{7}{l}{\textit{IPQ}} \\
IPQ spatial presence & 3.29 (0.82) & 3.98 (0.81) & 4.47 (0.84) & $.201$ & \boldmath $.003$\unboldmath & $.252$ \\
IPQ involvement & 3.05 (1.02) & 2.98 (0.88) & 3.45 (0.84) & $.203$ & $.446$ & $.926$ \\
IPQ realism & 2.80 (0.92) & 2.98 (0.85) & 4.07 (0.99) & \boldmath $.023$\unboldmath & \boldmath $.004$\unboldmath & $.754$ \\
\addlinespace[3pt]
\multicolumn{7}{l}{\textit{Self-authored experience}} \\
Q11.1 Inside the VE & 3.57 (1.09) & 4.36 (1.60) & 4.57 (1.02) & $1.000$ & \boldmath $.014$\unboldmath & $.235$ \\
Q11.2 Believability & 3.50 (1.22) & 3.36 (1.39) & 4.29 (1.14) & $.237$ & $.082$ & $1.000$ \\
Q11.3 Smoothness & 3.14 (1.41) & 3.79 (1.53) & 4.71 (0.83) & $.237$ & \boldmath $.007$\unboldmath & $.383$ \\
Q11.4 Task focus & 3.79 (1.19) & 4.64 (1.34) & 5.36 (0.93) & $.237$ & \boldmath $.003$\unboldmath & $.235$ \\
Q11.5 Felt immersed & 4.00 (1.36) & 4.50 (1.22) & 5.07 (0.83) & $.268$ & \boldmath $.037$\unboldmath & $.528$ \\
\addlinespace[3pt]
\multicolumn{7}{l}{\textit{Safety and risk items}} \\
Q11.6 Felt safe & 5.00 (0.88) & 3.21 (1.53) & 4.93 (0.92) & \boldmath $.003$\unboldmath & $.905$ & \boldmath $.012$\unboldmath \\
Q11.7 Obstacle cues & 4.43 (0.65) & 3.86 (1.23) & 5.50 (0.65) & \boldmath $<.001$\unboldmath & \boldmath $.002$\unboldmath & $.313$ \\
Q11.8 Recognize danger & 4.36 (0.74) & 3.79 (1.05) & 5.43 (0.65) & \boldmath $<.001$\unboldmath & \boldmath $.002$\unboldmath & $.194$ \\
Q11.9 Move confidently & 5.00 (1.18) & 3.79 (0.89) & 5.57 (0.85) & \boldmath $<.001$\unboldmath & $.187$ & \boldmath $.018$\unboldmath \\
Q11.10 Relative danger & 4.21 (0.89) & 3.21 (1.63) & 5.36 (0.50) & \boldmath $<.001$\unboldmath & \boldmath $.002$\unboldmath & $.124$ \\
Q11.11 Risk levels & 4.14 (0.77) & 3.07 (1.59) & 5.21 (0.70) & \boldmath $<.001$\unboldmath & \boldmath $.002$\unboldmath & $.124$ \\
Q11.12 Judgment confidence & 4.64 (1.01) & 4.21 (1.48) & 5.07 (0.83) & $.149$ & $.406$ & $.638$ \\
Q11.13 Misunderstanding & 2.14 (1.23) & 2.64 (1.95) & 2.43 (1.83) & $.924$ & $.905$ & $.792$ \\
\addlinespace[3pt]
\multicolumn{7}{l}{\textit{Symptoms}} \\
Q20.1 Dizziness & 1.82 (1.17) & 0.67 (1.15) & 0.38 (0.52) & $1.000$ & \boldmath $.014$\unboldmath & \boldmath $.033$\unboldmath \\
Q20.2 Nausea & 1.14 (0.69) & 0.22 (0.67) & 0.12 (0.35) & $1.000$ & \boldmath $.018$\unboldmath & \boldmath $.033$\unboldmath \\
Q20.3 Eyestrain & 1.56 (0.88) & 0.62 (1.19) & 0.57 (0.79) & $1.000$ & \boldmath $.043$\unboldmath & $.057$ \\
Q20.4 Sweating/heart rate & 1.57 (1.40) & 0.90 (1.29) & 0.43 (0.53) & $1.000$ & $.132$ & $.271$ \\
Q20.5 Disorientation & 1.22 (0.44) & 0.62 (0.92) & 0.55 (0.52) & $1.000$ & \boldmath $.019$\unboldmath & $.121$ \\
\addlinespace[3pt]
\multicolumn{7}{l}{\textit{Activity}} \\
Bubble touches & 120.07 (23.06) & 117.43 (32.23) & 126.50 (19.66) & $.137$ & $.566$ & $.883$ \\
Head proximity & 1.215 (0.046) & 1.203 (0.063) & 1.217 (0.063) & $.629$ & $1.000$ & $.462$ \\
Hand proximity & 0.814 (0.080) & 0.765 (0.074) & 0.799 (0.063) & $.340$ & $1.000$ & $.462$ \\
\bottomrule\end{tabular}
\Description{Rows report questionnaire scores, bubble touches, and head and hand proximity for Baseline, C1, and C2. Each row provides condition means and standard deviations followed by three adjusted pairwise comparisons. Bold values identify comparisons below the .05 threshold.}
\end{table*}

\FloatBarrier
\section{Texture Generation Prompt}\label{app:texture-generation-prompt}
We used the following prompt with \texttt{gpt-image-2} to generate textures. Each request included a screenshot of the virtual environment, with \texttt{\{assigned\_risk\_level\}} replaced by the object's assigned category: \texttt{low}, \texttt{medium}, or \texttt{high}.

\begin{quote}
\small
Generate a surface texture for a scanned physical object in virtual reality.

Use the attached screenshot to infer the virtual scene's theme, color palette, materials, and visual style. Identify scene-compatible materials or visual elements that suggest three levels of caution on contact: low risk through a soft or yielding appearance, medium risk through a firm or hard appearance requiring care, and high risk through an appearance suggesting that contact should be avoided.

The object's assigned risk level is \texttt{\{assigned\_risk\_level\}}. Generate one texture that communicates this level while fitting the scene. Use material appearance, color, and contrast to convey the intended degree of caution. Return only the texture image for application to the existing scanned mesh, without text labels or changes to the object's geometry.
\end{quote}
\FloatBarrier
\section{Study-Specific Questionnaire Items}\label{app:questionnaire}
Table~\ref{tab:questionnaire-items} lists the study-specific items administered after the VR task.

\begin{table*}[!htbp]
\caption{Study-specific questionnaire items, with identifiers matching those used in the main text.}
\label{tab:questionnaire-items}
\centering
\small
\renewcommand{\arraystretch}{1.15}
\begin{tabularx}{\textwidth}{@{}>{\raggedright\arraybackslash}p{0.10\textwidth}>{\raggedright\arraybackslash}X@{}}
\toprule
Item & Question wording \\
\midrule
\multicolumn{2}{@{}l}{\textit{Immersive experience}} \\
Q11.1 & During the VR experience, I felt like I was really `inside' the virtual environment. \\
Q11.2 & I felt that the virtual environment was realistic and believable. \\
Q11.3 & The visual presentation and interaction were smooth and natural. \\
Q11.4 & I was fully focused on the task without being distracted by the interface. \\
Q11.5 & Overall, I felt immersed in the virtual experience. \\
\midrule
\multicolumn{2}{@{}l}{\textit{Perceived safety}} \\
Q11.6 & I felt safe during the task. \\
Q11.7 & The system gave me enough cues to avoid obstacles safely. \\
Q11.8 & I could easily recognize potentially dangerous areas or objects. \\
Q11.9 & I was confident that I could move safely in the environment. \\
\midrule
\multicolumn{2}{@{}l}{\textit{Perceived risk differentiation}} \\
Q11.10 & The visual cues helped me understand which obstacles were more dangerous. \\
Q11.11 & I could distinguish different risk levels among obstacles. \\
Q11.12 & I was confident in my judgments about the risk level of obstacles. \\
Q11.13 & I sometimes misunderstood the level of danger presented by certain obstacles. \\
\midrule
\multicolumn{2}{@{}l}{\textit{Physical discomfort}} \\
Q20.1 & I experienced dizziness during the VR experience. \\
Q20.2 & I experienced nausea or upset stomach. \\
Q20.3 & I experienced eyestrain or visual fatigue. \\
Q20.4 & I experienced sweating or increased heart rate. \\
Q20.5 & I felt disoriented or lost balance at any point. \\
\bottomrule
\end{tabularx}
\Description{The table lists the exact wording and item identifiers of eighteen study-specific questionnaire items. Five items concern immersive experience, four concern perceived safety, four concern perceived risk differentiation, and five concern physical discomfort.}
\end{table*}
\end{document}